\documentclass[12pt]{article}

\def\pt{p_T}
\def\dis{distribution}
\def\bq{\begin{eqnarray}}
\def\eq{\end{eqnarray}}
\usepackage{graphicx}

\begin{document} 

\begin{center}  {\Large {\bf Ridge and Transverse Correlation at Separated Rapidities}}

\vskip .75cm
 {\bf Charles B.\ Chiu$^1$ and  Rudolph C. Hwa$^{2}$}

\vskip.5cm
{$^1$Center for Particles and Fields and Department of Physics,\\
University of Texas at Austin, Austin, TX 78712, USA\\
$^2$Institute of Theoretical Science and Department of
Physics,\\ University of Oregon, Eugene, OR 97403-5203, USA}
\end{center}

\vskip.5cm

\begin{abstract} 
 A simple phenomenological relationship between the ridge \dis\ in $\Delta\eta$ and the single-particle \dis\ in $\eta$ can be established from the PHOBOS data  on both \dis s. The implication points to the possibility that there is no long-range longitudinal correlation. An interpretation of the relationship  is then developed, based on the recognition that longitudinal uncertainty of the initial configuration allows for non-Hubble-like expansion at early time. It is shown that the main features of the ridge structure can be explained in a model where transverse correlation stimulated by semihard partons is the principal mechanism.
 
\end{abstract}
 
\vskip.5cm
\section{Introduction}
The ridge structure in two-particle correlation has been studied in nuclear collisions at the Relativistic Heavy-Ion Collider (RHIC) for several years \cite{ja, bia, ba, bia1, md} and is now also seen in $pp$ collisions at the Large Hadron Collider (LHC) \cite{cms}.  The nature of that structure is that it is narrow in $\Delta\phi$ (azimuthal angle $\phi$ relative to that of the trigger) but broad in $\Delta\eta$ (pseudorapidity $\eta$ relative to the trigger).  In Ref.\ \cite{ba} the range in $\Delta\eta$ is found to be as large as 4.  So far there is no consensus on the origin of the ridge formation \cite{int}.  It has been pointed out that the wide $\Delta\eta$ distribution implies long-range correlation \cite{ad,gmm}.  That is a view based partially on the  conventional estimate that the correlation length is about 2 \cite{kd}.  We make here a comparison between the $\eta$ ranges of single-particle distribution and two-particle correlation, using only the experimental data from PHOBOS \cite{ba,bb}. It is found that the large-$\Delta\eta$ ridge \dis\ is related simply to a shift of the inclusive \dis\ and an integral over the trigger $\eta$. That is a phenomenological observation without any theoretical input. Any successful model of ridge formation should be able to explain that relationship. 

There are subtleties about the single-particle \dis\ for all charges, $dN^{ch}/d\eta$, that to our knowledge has not been satisfactorily explained in all its details. Since it sums over all charges, hadrons of different types are included, making $dN^{ch}/d\eta$ to be quite different from $dN^{\pi}/dy$, which can be fitted by a Gaussian distribution in $y$ with width $\sigma_\pi=2.27$ \cite{bea}. That difference cannot be readily accounted for in any simple hadronization scheme. Fortunately, detailed examination of $dN^{ch}/d\eta$ is not required before we find its relationship to the ridge \dis\ $dN^{ch}_R/d\Delta\eta$, since both are for unidentified charged hadrons, and the empirical verification is based on the data from the same experimental group (PHOBOS).

As a consequence of the phenomenological relationship, we consider the possibility that there is no intrinsic long-range longitudinal correlation apart from what gives rise to the single-particle distribution. We have found that to generate $dN_R^{ch}/d\Delta\eta$ it is only necessary to have transverse correlation at different points in $\eta$, provided that at early time the small-$x$ partons do not expand in Hubble-like manner. If spatial uncertainty of wee partons are allowed at early time, the identification of spatial and momentum rapidities may not be valid near the tip of the forward light cone. Therein lies the origin of transverse correlation due to the possibility of near crossing of soft- and hard-parton trajectories. The energy lost by a hard parton  enhances the thermal energies of the medium partons in the vicinity of the hard parton's trajectory. The transverse broadening of any small-$x$ parton that passes through the cone of that enhancement leads to measurable effect of the ridge. The parton model that we use does not rely on flux tubes or hydrodynamics.

Recently, the existence of ridge has been called into question by investigations on the effect of fluctuations of the initial configurations in heavy-ion collisions \cite{ar,aglo}. Using hydrodynamical model and transport theory to relate the eccentricities of the  spatial initial state in the transverse plane to the azimuthal momentum anisotropy in the final state, it has been shown that the harmonic coefficients $v_n$ observed in the data can be understood in terms of such transverse fluctuations \cite{ty,xk,blo,ps,ral,qp,hm,sjg,gglo}. That is, however, only one of the possible interpretations of $v_n$. The effect of minijets on the initial configuration can yield similar consequences. In a companion article \cite{hz2} it is shown that the data on $v_n$ can also be well reproduced by taking the minijets into account in the recombination model without the details of hydrodynamics. Here, we raise the issue about the effect of longitudinal fluctuations that seem to be as important as transverse fluctuations, but have hardly been investigated.

After the phenomenological relationship between $dN^{ch}_R/d\Delta\eta$ and $dN^{ch}/d\eta$ is established in Sec.\ 2, we give our interpretation of the phenomenon in Sec.\ 3. It is not our objective to give a review of all other models that can reproduce the data on the ridge structure and assess their likelihood to explain the empirical observation made in Sec.\ 2.
We offer only to show the possibility  that the ridge can have the observed properties in the absence of long-range longitudinal correlation. Section 3 includes many subsections in which both longitudinal and transverse aspects of the correlation are examined in the parton model. Our conclusion is given in Sec.\ 4.

\section{Comparison between Ridge and Inclusive Distributions}

Our focus is on the PHOBOS data on two-particle correlation measured with a trigger particle having transverse momentum $p_T^{\rm trig} > 2.5$ GeV/C in Au + Au collisions at $\sqrt{s_{NN}} = 200$ GeV \cite{ba}.  The pseudorapidity acceptance of the trigger is $0 < \eta^{\rm trig} < 1.5$.  The per-trigger ridge yield integrated over $|\Delta\phi| < 1$, denoted by $(1/N^{\rm trig})dN_R^{ch}/d\Delta\eta$, includes all charged hadrons with $p_T^a\ ^>_\sim\ 7$ MeV/c at $\eta^a = 3$ and $p_T^a\ ^>_\sim\ 35$ MeV/c at $\eta^a = 0$, where the superscript $a$ stands for associated particle in the ridge.  For simplicity we use the notation $\eta^{\rm trig} = \eta_1$, $\eta^a = \eta_2$, $\Delta\eta = \eta_2 - \eta_1$, $\phi^{\rm trig} = \phi_1$, $\phi^a = \phi_2$, $\Delta\phi = \phi_2 - \phi_1$.  Since all ridge particles are included in the range $|\Delta\phi| < 1$,  the $\Delta\phi$ dependence of the ridge structure does not show up in the properties of $dN_R^{ch}/d\Delta\eta$. We have previously studied the $\Delta\phi$ dependence of the ridge \cite{17}, which will be summarized below in Sec.\ 3.2. Here we focus on   our aim  to relate the ridge distribution in $\Delta\eta$ to the single-particle distribution in $\eta$.  We first make a phenomenological observation using only PHOBOS data for both distributions.  After showing their relationship, we then make an interpretation that does not involve extensive modeling.

To do meaningful comparison, it is important to use single-particle $\eta$ distribution, $dN^{ch}/d\eta$, that has the same kinematical constraints as the ridge distribution.  That is, it involves an  integration over $p_T$ and a sum over all charged hadrons
\begin{eqnarray}
{dN^{ch}\over d\eta} = \sum_h \int dp_Tp_T\rho_1^h (\eta,p_T)  ,    \label{1}
\end{eqnarray}
where $\rho_1^h(\eta,\pt)=dN^h/\pt d\pt d\eta$, and the lower limit of the $p_T$ integration is $35(1-\eta/3.75)$ MeV/c in keeping with the acceptance window of $p_T^a$ \cite{ba}.  The data on $(1/N^{\rm trig})dN_R^{ch}/d\Delta\eta$ are for 0-30\% centrality.  PHOBOS has the appropriate $dN^{ch}/d\eta$ for 0-6\%, 6-15\%, 15-25\% and 25-35\% centralities \cite{bb}, as shown in Fig.\ 1(a). Thus we average them over those four bins.  The result is shown in Fig.\ 1(b) by the small circles for 0-30\% centrality.  Those points are fitted by the three Gaussian \dis s, located at $\eta=0$ and $\pm \hat\eta$,
\begin{eqnarray}
{dN^{ch}\over d\eta} = A \{\exp[-\eta^2/{2\sigma_0^2}] + a_1\exp[-(\eta-\hat\eta)^2/{2\sigma_1^2}] + a_1\exp[-(\eta+\hat\eta)^2/{2\sigma_1^2}]\}  \label{2}
\end{eqnarray}
shown by the solid (red) line in that figure with $A=468, \sigma_0=2.69,  a_1=0.31, \hat\eta = 2.43, \sigma_1=1.15$.  The dashed line shows the central Gaussian, while the dash-dotted line shows the two side Gaussians.
The purpose of the fit is mainly to give an analytic representation of $dN^{ch}/d\eta$ to be used for comparison with the ridge distribution.  
Nevertheless, it is useful to point out that the width $\sigma_0$ of the central Gaussian in $\eta$ is larger than the width of the pion $y$-distribution, $\sigma_\pi=2.27$, mentioned in Sec.\ 1. The two side-Gaussians are undoubtedly related to the production of protons, since BRAHMS data show significant $p/\pi$ ratio above $\eta=2$ and $p_T>1$ GeV/c \cite{iga}. The value of $\hat\eta$ in Eq.\ (\ref{2}) being $>2$ is a result of the enhancement by proton production. Any treatment of correlation among charged particles without giving proper attention to the protons is not likely to reproduce the inclusive distribution given by  Eq.\ (\ref{2}), whose $\eta$ width is significantly stretched by the side-Gausssians.

\begin{figure}[tbph]
\vspace*{.5cm}
\includegraphics[width=.9\textwidth,clip]{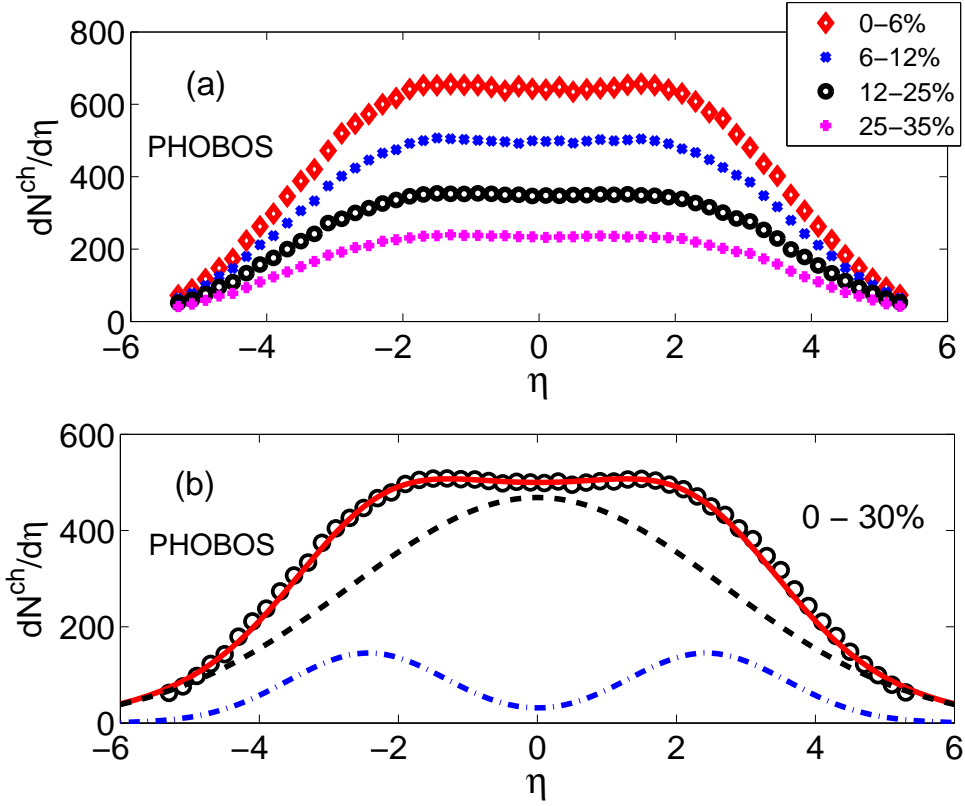}
\caption{(Color online) Pseudorapidity \dis\ in Au-Au collisions at $\sqrt{s_{NN}}=200$ GeV for (a) various centrality bins  and (b) 0-30\% centrality. Data are from Ref.\ \cite{bb}. The (red) line in (b) is a fit using Eq.\ (\ref{2}), whose first term is represented by the  dashed line and the other two terms by the dash-dotted line.}
\end{figure}

We now propose the formula
\begin{eqnarray}
{1\over N^{\rm trig}} {dN_R^{ch}\over d\Delta\eta} =r\int_0^{1.5} d\eta_1  \left. {dN^{ch}\over d\eta_2} \right|_{\eta_2 = \eta_1 + \Delta\eta},     \label{3}
\end{eqnarray}
where $r$ is a parameter that summarizes all the experimental conditions that lead to the magnitude of the ridge distribution measured relative to the single-particle distribution. In particular, $r$ does not depend on $\eta_1$ or $\eta_2$; otherwise, the equation is meaningless in comparing the $\eta$ dependencies.

There is no theoretical input in Eq.\ (\ref{3}), except for the question behind the proposal:  how much of the $\Delta\eta$ distribution can be accounted for by just a mapping of $dN^{ch}/d\eta_2$ with a shift due to the definition $\Delta\eta = \eta_2 - \eta_1$, and a smearing due to the trigger acceptance, $0 < \eta_1 < 1.5$?  Another way of asking the question is: how would the range of correlation be affected if the experimental statistics were high enough so that the trigger's $\eta$ range can be very narrow around $\eta_1 = 0$?

\begin{figure}[tbph]
\includegraphics[width=1\textwidth,clip]{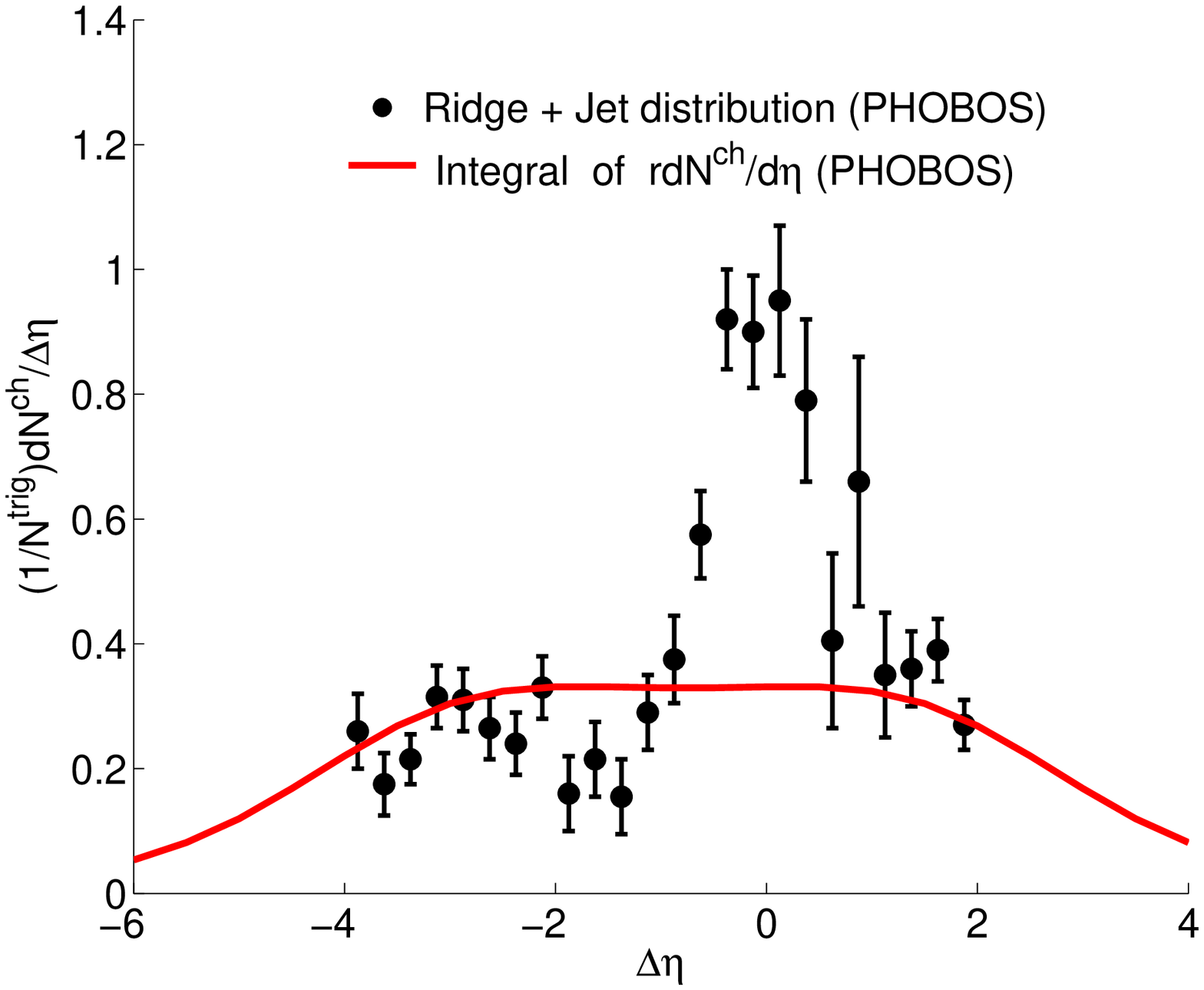}
\vspace*{-1.2cm}
\caption{(Color online) Two-particle correlation of charged particles. Data are from Ref.\ \cite{ba} that include both ridge and jet components. The line is  a plot according to Eq.\ (\ref{3}) using $\eta$ \dis\ from Fig.\ 1 \cite{bb}.}
\end{figure}

The proposed formula in Eq.\ (\ref{3}) is tested by substituting the fit of $dN^{ch}/d\eta$ according to Eq.\ (\ref{2}) into the integrand on the right-hand side.  The result is shown in Fig.\ 2 with $r$ being adjusted to fit the height of the ridge distribution; its value is $4.4\times 10^{-4}$. The peak in the data around $\Delta\eta=0$ is, of course, due to the jet component associated with the trigger jet and is not relevant to our comparison here. 
That component has been studied in the recombination model as a consequence of thermal-shower recombination that can give a good description of the peak both in $\Delta\eta$ and $\Delta\phi$ \cite{12}. For the ridge considered here, it is evident that the large $\Delta\eta$ distribution in Fig.\ 2 is well reproduced by Eq.\ (\ref{3}). Since our concern is to elucidate the implications of the range of $\Delta\eta$, we leave the fluctuation from the flat distribution in the interval $-2<\Delta\eta<-1$ as an experimental problem.
 In qualitative terms the width of the ridge distribution is due partly to the width of $dN^{ch}/d\eta$ and partly to the smearing of $\eta_1$, which adds another 1.5 to the width.  No intrinsic dynamics of long-range longitudinal correlation has been put in.  Note that the center of the plateau in $\Delta\eta$ is at $-0.75$, which is the average of the shift due to $\eta_1$ being integrated from 0 to 1.5. It suggests that if $\eta_1$ were fixed at $\eta_1\approx 0$ when abundant data become available, then the width of $dN_R^{ch}/d\Delta\eta$ would be only as wide as that of the single-particle $dN^{ch}/d\eta$. No theoretical prejudice has influenced these observations.

\section{Interpretation of Phenomenological Observation}

	We now consider an interpretation of what Eq. (\ref{3}) implies, given the empirical support for its validity from Fig.\ 2. First, we ask what the implication of the phenomenological observation is in terms of the range of longitudinal correlation. Then we describe a model for ridge formation first for azimuthal dependence at mid-rapidity, then for larger pseudo-rapidity pertinent to the data. The considerations from various perspectives lead to the notion of transverse correlation that will become the core element of our model to explain the ridge phenomenon.

\subsection{Range of Longitudinal Correlation}

 Since the observed ridge \dis\ integrates over trigger $\eta$, we write it as
\bq
{1\over N^{\rm trig}}{dN_R^{ch}\over d\Delta\eta}=\int_0^{1.5} d\eta_1\sum_{h_2} \int dp_2 p_2\left. R^{h_2}(\eta_1,\eta_2,p_2) \right|_{\eta_2 = \eta_1 + \Delta\eta},     \label{4}
\eq
where we exhibit also explicitly the sum over the hadron type of the ridge particle $h_2$ and the integral over its transverse momentum, denoted by $p_2$. According to the definition of correlation $C_2(1,2)=\rho_2(1,2)-\rho_1(1)\rho_1(2)$, we can express the per-trigger ridge correlation as
\bq
R^{h_2}(\eta_1,\eta_2,p_2) = \sum_{h_1} \int dp_1 p_1 {\rho_2^{h_1h_2{(B+R)}}(\eta_1,p_1,\eta_2,p_2)\over \rho_1^{h_1}(\eta_1,p_1)} -\rho_1^{h_2{(B)}}(\eta_2,p_2),  \label{5}
\eq
where $p_1$ is the transverse momentum of the trigger particle; $B$ and $R$ in the superscript denote background and ridge, respectively. The jet component
in the associated-particle \dis\ is excluded in Eq.\ (\ref{5}).

On the other hand, with Eq.\ (\ref{1}) substituted into Eq.\ (\ref{3}) we  have, using $\eta_2$ and $p_2$ instead of 
$\eta$ and $p_T$,
\bq
{1\over N^{\rm trig}}{dN_R^{ch}\over d\Delta\eta}=\int_0^{1.5} d\eta_1\sum_{h_2} \int dp_2 p_2\left. r \rho_1^{h_2}(\eta_2,p_2) \right|_{\eta_2 = \eta_1 + \Delta\eta}.     \label{6}
\eq
Comparing Eq.\ (\ref{6}) to (\ref{4}) we see that the ridge \dis\ $R^{h_2}(\eta_1,\eta_2,p_2)$ is to be related to the phenomenological quantity $r\rho_1^{h_2}(\eta_2,p_2)$. Thus the crux of the relationship between the ridge and inclusive \dis s involves the interpretation of $r\rho_1^{h_2}$. To that end let us first write $\rho_1^{h_2}$ in the form
\bq
\rho_1^{h_2}(\eta_2,p_2)={dN^{h_2}\over d\eta_2p_2dp_2}=H^{h_2}(\eta_2,p_2) V(p_2), 
\qquad V(p_2)=e^{-p_2/T}, \label{7}
\eq
where $V(p_2)$ is the transverse component that contains the explicit exponential behavior of $p_2$. Although $H^{h_2}(\eta_2,p_2)$ has some mild $p_2$ dependence due mainly to mass effects of $h_2$, the average transverse momentum $\left<p_2\right>$ is determined primarily by the inverse slope $T$ and is not dependent on $\eta_2$. This is an approximate statement that is based on the BRAHMS data \cite{bea}, which show that $\left<p_T\right>$ is essentially independent of rapidity. Since $r$ serves as the phenomenological bridge between $R^{h_2}$ and $\rho_1^{h_2}$, the key question to address is: which of the two components, the longitudinal $H^{h_2}(\eta_2,p_2)$ or the transverse $V(p_2)$, does the two-particle correlation 
generated by a trigger at $\eta_1$ 
exert its most important influence in relating $R^{h_2}$ to $\rho_1^{h_2}$?

If there is longitudinal correlation from early times as in \cite{ad,gmm,10}, then its effect must be to convert $H^{h_2}(\eta_2,p_2)$ to $R^{h_2}(\eta_1,\eta_2,p_2)$. In that case $V(p_2)$ is relegated to the secondary role due to radial flow (which is, nevertheless, essential in explaining the $\Delta\phi$ restriction as in Refs.\ \cite{gmm,sv,shu}). On the other hand, if there is no intrinsic long-range longitudinal correlation, then $H^{h_2}(\eta_2,p_2)$ is unaffected, and the ridge can only arise from the change in the transverse component, $V(p_2)$, due to a hard scattering that leads to the trigger. Without phenomenology one would think that the first option is more reasonable, when $|\Delta\eta|\sim 4$ is regarded as large, and especially when there is an inclination based on theoretical ideas that there is long-range correlation. With the ridge phenomenology described by Eq.\ (\ref{3}) pointing to direct relevance of $H^{h_2}(\eta_2,p_2)$, the question becomes that of asking:   $|\Delta\eta|$ is large compared to what? If it is now recognized that $|\Delta\eta|$ is not large compared to the $\eta_2$ range of $\rho_1^{h_2}(\eta_2,p_2)$ after the widening due to $\eta_1$ smearing (remarked at the end of the previous section) is taken into account,
then the need for a long-range dynamical correlation to account for the structure of $R^{h_2}(\eta_1,\eta_2,p_2)$ is  lost. We describe below a possible explanation based on the second option of no long-range correlation. The key is to accept the suggestion of the data that the unmodified  longitudinal component $H^{h_2}(\eta_2,p_2)$ is sufficient.

A series of articles have treated the subject of ridge formation in the recombination model \cite{hy}, beginning with (a) the early observation of pedestal in jet correlation \cite{12,11}, to (b) its effects on azimuthal anisotropy of single-particle \dis\ at mid-rapidity \cite{13,14}, and then to (c) the dependence on the azimuthal angle $\phi_s$ of the trigger relative to the reaction plane \cite{15,17,16,18}. Forward productions in d-Au and Au-Au collisions have also been studied in \cite{19,20}. Our consideration here of ridge formation at $|\Delta\eta|>2$ is an extension of earlier studies with the common theme that ridges are formed as a consequence of energy loss by semihard or hard partons as they traverse the medium. The details involve careful treatment of the hadronization process with attention given to both the longitudinal and transverse components. The $\phi$ dependence has been studied thoroughly in \cite{17,18}, and the $\eta$ dependence should take into account of
 the experimental fact that the $p/\pi$ ratio can be large $(> 2.5)$ at large $\eta$ \cite{iga} so that $H^{h_2}(\eta_2,p_2)$  in Eq.\ (\ref{7}) can be properly reproduced.
  
 \subsection{Azimuthal Dependence of the Ridge}
 
 We give in this subsection a brief summary of the $\Delta\phi$ \dis\ that we have obtained previously in our treatment of the ridge formation \cite{17}. In so doing we also explain more thoroughly an aspect of the basic elements of our model.
 
 The tenets of our interpretation of the ridge structure are that its formation is due to (a) the passage of a semihard parton through the medium, and (b) the conversion of the energy loss by the parton to the thermal energy of the soft partons in the vicinity of its trajectory. Hadronization of the enhanced thermal partons at small rapidity forms the ridge standing above the background. In Ref.\ \cite{17} we have considered the geometry of the trajectory of a semihard parton traversing the medium in the transverse plane at mid-rapidity, $|\eta|<1$, taking into account the azimuthal angle $\phi_s$ of the trajectory that is to be identified with the trigger direction relative to the reaction plane. Along that trajectory, labeled by points $(x,y)$ in the transverse plane, the medium expands in the direction $\psi(x,y)$. If $\psi(x,y)$ is approximately equal to $\phi_s$ for most of the points $(x,y)$ along the trajectory of the semihard parton, then the thermal partons enhanced by successive soft emissions are carried by the flow along in the same direction; the effects reinforce one another and lead to the formation of a ridge in a narrow cone. On the other hand, if the two directions are orthogonal, then the soft partons  emitted from the various points along the trajectory are dispersed over a range of surface area, so their hadronization leads to no pronounced effect. These  extreme possibilities suggest a correlation function between $\phi_s$ and $\psi$, which we assume to have the Gaussian form
\begin{eqnarray}
C(x, y, \phi_s) = \exp \left[ -{(\phi_s-\psi(x, y))^2\over 2 \lambda}\right] \ ,  
\label{7a}
\end{eqnarray}
where the width-squared $\lambda$ is a parameter to be determined. This correlation is the central element of our Correlated Emission Model (CEM) \cite{17}. 

Considerable care is given to the calculation of the observed ridge yield $Y(\phi_s)$ as a function of $\phi_s$. It involves integrations over the path length of the trajectory of the semihard parton and its point of creation in the medium whose density depends on nuclear overlap, etc. To compare with the data on $Y(\phi_s)$ we also have to integrate over all $\phi$ of the ridge particle. It is found that by adjusting the value of $\lambda$ it is possible to fit the data on $Y(\phi_s)$ in the entire range $0<\phi_s<\pi/2$ for both 0-5\% and 20-60\% centralities. The value determined is $\lambda=0.11$, corresponding to a width $\sigma_c=\sqrt\lambda=0.34$ rad, which is much smaller than the width of the ridge itself, $\Delta\phi\sim 1$. We have been able to show that using $\lambda=0.11$ the calculated \dis\ of the ridge $dN_R/\Delta\phi$ agrees well with the data. We further made a prediction on the existence of an asymmetry property of the ridge $R(\phi,\phi_s)$ in its $\phi$ dependence relative to $\phi_s$. That prediction was subsequently verified by the STAR data \cite{pkn}.

The mechanism for $\phi$ correlation described above will form the basis of transverse correlation when we move away from mid-rapidity to $|\eta|>1$. It is necessary, however, to start the consideration with a discussion of the forward-moving soft partons relative to the semihard partons at early time.

\subsection{Longitudinal Initial Configuration}

We now extend the mechanism for ridge formation at mid-rapidity described above to $|\eta|>1$. Of course, without examining $dN_R/d\Delta\eta d\Delta\phi$ at $|\Delta\eta|>1$ one cannot strictly refer to the structure at $|\Delta\eta|<1$ as ridge, which by definition has a flat \dis\ in $\Delta\eta$, but is restricted in $|\Delta\phi|$. We have actually considered the $\Delta\eta$ behavior before we investigated the $\Delta\phi$ structure at a time when the ridge was referred to as pedestal \cite{12}. Calculation was done in the framework where the trigger is formed by thermal-shower recombination and the associated particles in the ridge by the recombination of enhanced thermal partons. In view of our present phenomenological finding in Fig.\ 2 and expressed in Eqs.\ (\ref{3}) and (\ref{6}), we reformulate our model here with attention given to the initial configuration relevant to the problem at hand.

In the preceding subsection we have discussed the correlation between the semihard parton at $\phi_s$ and the local flow direction at $\psi(x,y)$, expressed in Eq.\ (\ref{7a}) for $|\eta|<1$.  To extend the same mechanism to $|\eta|>1$, it is important to recognize first that the longitudinal momenta of the hadrons produced outside the mid-rapidity region are not generated by the semihard parton, as it would be ruled out simply by energy conservation. In accordance to the original parton model \cite{rf}, the right- and left-moving partons in the initial configuration provide the main thrust for forward and backward momenta. To be more quantitatively pertinent to the ridge structure observed in \cite{ba}, let us recall that the pseudo-rapidity ranges of the trigger and ridge particles are $0<\eta^{\rm trig}<1.5$ and $-4<\Delta\eta<2$. For the sake of discussing positive momentum fractions, let us reverse the signs of $\eta$ without loss of generality, and regard $\eta_1>-1.5$ and $\eta_2<2.5$ so that $-2<\eta_2-\eta_1<4$. Let us  be  generous and set $\eta_2<3$; it corresponds to $\theta_2>0.1$. That is, a ridge particle has $p_T/p_L=\tan\theta_2 > 0.1$. Assuming an average $\left<p_T\right>\sim 0.4$ GeV/c implies $p_L < 4$ GeV/c. The coalescing quarks that form a pion at such a $p_L$ would have on average a longitudinal momentum of $k_L<2$ GeV/c (even less for a proton). For $\sqrt{s_{NN}}=200$ GeV, the corresponding momentum fraction $x$ of the quarks is $2k_L/\sqrt s<0.02$. That is not a large-$x$ soft parton, being very nearly in the wee region \cite{rf}. Thus the kinematics of the particles in the ridge does not indicate that the coalescing quarks are very much in the forward (or backward) fragmentation region.

For $\sqrt s/2=100$ GeV the Lorentz contraction factor is sometimes taken to be $\gamma\sim 100$, but that corresponds to $x=1$, where no quarks exist. If we take the average valence-quark momentum  fraction to be $\left<x_{\rm val}\right>\sim 1/4$, then the corresponding $\gamma$ is $\sim 25$ and $\Delta z_{\rm val}\sim 2R_A/\gamma\sim 0.5$ fm, which has a width that is not  very thin. When two such slabs overlap in the initial configuration, the wee partons of the Au-Au colliding system can occupy a wider longitudinal space ($\Delta z\sim 2$ fm) of uncertainty due to quantum fluctuations  --- 1 fm on each side of the overlapping slabs consisting of soft parton with $x$ much smaller than $\left<x_{\rm val}\right>$. Our point is then that in that space of $\Delta z\sim 2$ fm in the initial configuration quantum fluctuations free us from requiring the soft partons to follow a Hubble-like expansion, i.e., the faster partons are on the outer edges of that longitudinal space, right-moving ones on the right side, and left-moving ones on the left. Note that we have this freedom because we have not restricted ourselves to a dynamical picture of flux tube being stretched by receding thin disks, as in \cite{gmm,10}.

\begin{figure}[tbph]
\vspace*{-1cm}
\hspace*{-2cm}
\rotatebox{90}{\includegraphics[width=1\textwidth,clip]{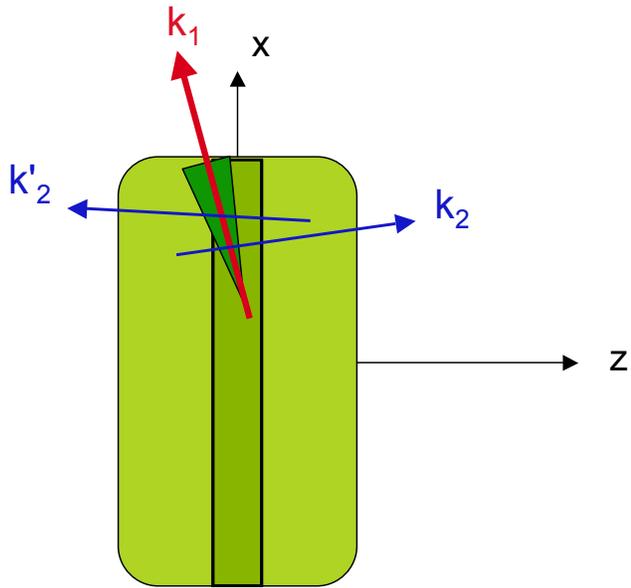}}
\vspace*{-4cm}
\caption{(Color online) A sketch of initial configuration in $x$-$z$ plane at early time. Horizontal thickness of the medium is $\Delta z\sim 2$ fm; the inner vertical slab indicates the relative thickness ($\sim 0.5$ fm) of the overlapping contracted disks in which the valence quarks are restricted. Red arrow represents semihard parton surrounded in-medium by a cone of enhanced thermal partons. Blue arrows represent soft partons with $k_i\ ^<_\sim \ 2$ GeV/c that originate from outside the slab and can therefore interact with the cone.}
\end{figure}

For a trigger particle to have $p_T^{\rm trig}>2.5$ GeV/c the initiating semihard or hard parton must have $k_T>3$ GeV/c and is created at early time. In Fig.\ 3 we show a sketch of the initial configuration in $x$-$z$ plane that depicts the relationship among  various possible momentum vectors at that time. The horizontal thickness of the shaded region is $\Delta z\sim 2$ fm and the vertical height is $2R_A\sim 12$ fm, thus not to scale. The central slab marked by a darker region of $\Delta z_{\rm val}\sim 0.5$ fm represents the longitudinal extent in which the  valence quarks are contracted. The (red) arrow labeled $k_1$ is the semihard parton that initiates the trigger; it starts from inside the narrow slab because the longitudinal momenta of the colliding partons before scattering are high. The two other (blue) arrows labeled $k_2$ and $k'_2$ represent two possible soft partons with $k_L\ ^<_\sim\ 2$ GeV/c, originating from outside the inner slab, since their $\Delta z$ is larger than $\Delta z_{\rm val}$. We place those vectors in such positions to emphasize the possibility that they can originate from the opposite sides of the slab. 
That is what we mean by  expansion at early time that is not of Hubble-type. The conical region (shaded green) around vector $k_1$ represents the vicinity of the trajectory of the semihard parton where the thermal partons are enhanced due to the energy loss by the semihard parton. Note that since the soft partons $k_2$ and $k'_2$ have larger $\Delta z$ than that of the valence quarks, they can cross the conical region, so the transverse components of the soft partons can be broadened by their interaction with the enhanced thermal partons.

\subsection{Transverse Correlation}

The discussion above on the space-momentum relationship between the semihard and soft partons at early time in the uncertainty region $\Delta z$ gives the conceptual basis for our view of how hadrons in the ridge are formed at late time. Our main point about the initial longitudinal uncertainty is that the forward-moving soft partons that eventually hadronize can be influenced by the semihard parton because the soft-parton trajectory starting from the left side of the central slab shown in Fig.\ 3 can traverse the cone of enhanced thermal partons. To be more quantitative we return to the general factorizable form of the single-particle \dis\ given in Eq.\ (\ref{7}) where $p_2$ refers to the transverse component $p_T$ of particle 2. The effect of the semihard parton on particle 2 is the transverse broadening of the soft parton $k_2$ in Fig.\ 3, in much the same way that the Cronin effect is conventionally explained in terms of initial-state broadening \cite{22}. That is, $V(p_2)$ in Eq.\ (\ref{7}) is modified if (a) there is a semihard parton $k_1$, and (b) $k_2$ (and other soft partons not shown in Fig.\ 3) passes through the cone in the vicinity of $k_1$. We denote the case without the semihard parton by $V_B(p_2)$ representing the background, where
\bq
V_B(p_2)=\exp(-p_2/T_0),  \label{9a}  
\eq
and the case with semihard parton and with $\phi$ in the vicinity of the cone by
\bq
V_{B+R}(p_2)=\exp(-p_2/T),  \label{10a}  
\eq
where $T>T_0$ is a result of the interaction with the enhanced thermal partons. Then the ridge has a transverse component that rises above the background and has the $\pt$ dependence
\bq
V_R(p_2)=V_{B+R}(p_2)-V_B(p_2).   \label{11a}  
\eq
This is the essence of transverse broadening due to the presence of semihard parton.
Since  the soft partons $k_2$ must pass through the enhanced cone (narrow in $\phi$) in order to develop transverse broadening, they contribute to the ridge only within the $\Delta\phi$ interval around $\phi_1$, discussed in Ref.\ \cite{17}.

The transverse correlation that we discuss here is not what one usually associates with the correlation between hadrons in the fragments of a high-$\pt$ jet. All of those fragments are in a small range of $\Delta\eta$ and have transverse-momentum fractions that are correlated. They populate the peak in Fig.\ 2. In our problem about the ridge we have been concerned with the transverse momentum of a particle associated with a trigger outside that peak. The former reveals the effect of the medium on the  jet, while the latter reveals the effect of the jet on the medium. That is the basic difference between the jet and ridge components of the associated particles. Since semihard or hard scattering takes place early, transverse broadening can take place for soft partons (the medium) moving through the interaction zone, leading to the ridge structure.

\subsection{The Ridge}

We may now write the per-trigger ridge correlation \dis\ $R^{h_2}(\eta_1,\eta_2,p_2)$ that is introduced in Eqs.\ (\ref{4}) and (\ref{5}) in the form
\bq
R^{h_2}(\eta_1,\eta_2,p_2)=c H^{h_2}(\eta_2,p_2) V_R(\Delta\eta,p_2),  \label{12a}
\eq
where, for $\Delta\eta$ in the range of the ridge, $V_R(\Delta\eta,p_2)$ may be approximated by $V_R(p_2)$ given in Eq.\ (\ref{11a}), i.e., 
\bq
V_R(p_2) =e^{-p_2/T}-e^{-p_2/T_0}= e^{-p_2/T}(1-e^{-p_2/T'}) ,  \qquad\quad  T'={T_0T\over T-T_0}.   \label{12b}
\eq
As we have seen in Fig.\ 2 and Eq.\ (\ref{3}), that range of $\Delta\eta$ where $T>T_0$ is no more than the $\eta_2$ range of $dN^{ch}/d\eta_2$, which in turn is determined by the $\eta_2$ range of $H^{h_2}(\eta_2,p_2)$ in Eq.\ (\ref{12a}). Thus in practice we may suppress the $\Delta\eta$ dependence in $V_R(\Delta\eta,p_2)$. The constant $c$ in Eq.\ (\ref{12a})  characterizes the magnitude of the ridge, which can depend  on many factors that  include the fluctuations in the initial configuration, the details of correlation dynamics, the experimental cuts, the
$\Delta\phi$ interval where the ridge is formed and the related scheme of background subtraction. Its value (that was not calculated) does not affect the relationship between the $\eta$ dependencies of the two sides of Eq.\ (\ref{12a}).

The expression for $V_R(p_2)$ in Eq.\ (\ref{12b}) was first obtained in Refs.\ \cite{13,14} as a description of the ridge \dis\ without trigger. It was noted there that $V_R(p_T)\to 0$ as $p_T\to 0$, and that $p_T/T'$ sets the scale for $v_2(p_T,b)$ for $p_T<0.5$ GeV/c in agreement with the data on it. More recently, a detailed study of $v_2(p_T,b)$ and the inclusive distribution has been carried out in Ref.\ \cite{hz2}, where it is found that  $T_0=0.245$ GeV and $T=0.283$ GeV, so that $T'=1.825$ GeV. The exact values are not important to our qualitative conclusion to be drawn below.

To proceed, we now substitute Eq.\ (\ref{12a}) in (\ref{4}) and use (\ref{7}) to obtain
\bq
{1\over N^{\rm trig}}{dN_R^{ch}\over d\Delta\eta}=\int_0^{1.5} d\eta_1\sum_{h_2} \int dp_2 p_2\left. {cV_R(p_2)\over V(p_2)} \rho_1^{h_2}(\eta_2,p_2) \right|_{\eta_2 = \eta_1 + \Delta\eta}.     \label{13}
\eq
Comparing this equation with Eq.\ (\ref{6}), we come to the conclusion that $r$ is a phenomenological approximation of  $c V_R(p_2) / V(p_2)$ in the region where it contributes most to the integral over $p_2$. From Eq.\ (\ref{12b}) we get $ V_R(p_2) / V(p_2)=1-e^{-p_2/T'}$ which is  severely damped by the exponential decrease of $\rho_1^{h_2}(\eta_2,p_2)$ in Eq.\ (\ref{13}) for $p_2>1$ GeV/c, since $T'\gg T$. Thus $c V_R(p_2) / V(p_2)$ may be approximated by a constant $r$ in the region where the integrand is maximum at around $p_2\sim 0.5$ GeV/c. In so doing, we obtain Eq.\ (\ref{6}) and therefore the phenomenological relation given by Eq.\ (\ref{3}).

Equation (\ref{13}) implies that there is transverse correlation, but no explicit longitudinal correlation beyond what is implicitly contained in $\rho_1^{h_2}$. 
In words, we can summarize the characteristics of the ridge as being generated by the same dynamical mechanism at any $\eta$ in the range where single-particle distribution can reach. That mechanism depends on semihard or hard partons (with or without trigger) whose energy loss to the medium leads to transverse broadening of small-$x$ partons that encounter the enhanced region of thermal partons. The transverse-momentum distribution of the ridge particles is the same for any $\eta$, and the $\eta$ range of the ridge is no more than that of the single-particle inclusive distribution because the partonic origin of the longitudinal momentum of any particle is the same.

Recently, an extension of similar consideration as described here has been applied to the study of ridge formation in $pp$ collisions at LHC and succeeded in explaining the $p_T$ dependence of the ridge yield found by the CMS Collaboration at $\sqrt s=7$ TeV  \cite{vk,29}.

\section{Conclusion}

An issue that this study has brought up is the usage of the word ``large'' in reference to the  range of $\Delta\eta$ in the ridge structure. Our  phenomenological observation in Eq.\ (\ref{3}), substantiated by Figs.\ 1 and 2, does not reveal any quantitative definition of what large $\Delta\eta$ means. 
 To be able to relate large $\Delta\eta$ to dynamical long-range correlation is a worthy theoretical endeavor, but more can be added to its phenomenological relevance if it can also elucidate the empirical connection between the two sides of Eq.\ (\ref{3}).

The approach that we have taken  involves no long-range longitudinal correlation for the ridge. The observed ridge \dis\ is interpreted in our approach as being due to transverse correlation with a range in $\Delta\eta$ that is no more than that of the single-particle \dis. That is, the $p_T$ \dis s of the detected hadrons in the ridge have a larger inverse slope than that of the particles outside, which have larger $\Delta\phi$ than the ridge width. We have described the partonic basis for how the transverse correlation at different $\eta$ depends on longitudinal fluctuations in the initial state. Furthermore, without hard partons there can be no ridge associated with a trigger. In the absence of a trigger there is still a ridge component hidden in the single-particle distribution due to semihard partons that are more copiously produced than hard partons. The $\pt$ \dis s of the ridge particles are essentially the same \cite{14,18,hz2}.

If a hard (or semihard) scattering is likened to an earthquake, then the ridge is the counterpart of tsunami, and the thermal medium carrying the enhancement is the ocean water. Transverse correlation is the rise in water level at various points along a coast hit by the tsunami. Although the tsunami damage is insensitive to the horizontal separation among the coastal cities, it  should not be interpreted as evidence for long-range horizontal (longitudinal) correlation. The buildings in different cities are not horizontally correlated, but their uprooting by vertical displacements is a sign of transverse correlation caused by the tsunami.
Similarly, there is transverse correlation at various points in the ridge but no long-range longitudinal correlation.  Where the analogy fails, as all analogies do at some point, is that our expanding system illustrated in Fig.\ 3 is not Hubble-like in the initial configuration and that the soft partons must intersect the enhanced cone of the hard parton in order to carry the effect of enhancement at $|\Delta\eta|>1$. That is where the restriction in $\Delta\phi$ enters in the ridge  problem. 
There is no such complication in the earthquake/tsunami example, which is strictly a classical case of wave propagation.
Another point where the analogy may be misleading is that in the case of the tsunami the energy of wave propagation is provided entirely by the earthquake. In our problem the momenta of the  forward-moving soft partons are in the initial state whether or not there is a hard (or semihard) scattering. They are the medium; their transverse momenta can be enhanced to form a ridge in the same way that the ocean water can be perturbed by the earthquake to develop a tsunami, 
whose underlying medium, however, does not expand. 
 Note that in both cases the detection of trigger or earthquake is not essential in assessing the effect of ridge or tsunami. The main point of the analogy is to illustrate the meaning of transverse correlation at separated rapidities without longitudinal correlation (and without suggesting similarity in dynamics).
 
 A crucial point in our interpretation of the ridge phenomenon is that the quantum fluctuation of the longitudinal coordinates of the initial configuration is important, as illustrated in Fig.\ 3. Because of the possibility that low-$x$ partons with positive momenta do not necessarily have to be located on the positive side of the thinner slab to which the high-$x$ partons are contracted, the usual approximation that equates spatial rapidity with momentum rapidity should not be extended to the neighborhood of the tip of the forward light cone. Fluctuations of the initial longitudinal configuration are not usually considered. In contrast, fluctuations of the initial transverse configuration have been investigated vigorously in recent years, leading to results according to hydrodynamical expansion that have significant phenomenological consequences on the transverse structure quantified by the 
 azimuthal harmonics, one of which being the diminution of the ridge itself. Here we find that longitudinal fluctuation of the initial parton configuration can be the source of the longitudinal structure in the ridge phenomenon. More detailed work is obviously needed to put on firmer footing the ideas initiated here.

To sum up, we have two  important  findings to emphasize.  One is the phenomenological  relationship between $dN_R^{ch}/d\Delta\eta$ and $dN^{ch}/d\eta$. The other is an interpretation of that relationship in terms of transverse correlation without long-range longitudinal correlation.

\section*{Acknowledgment}
This work was supported, in part,  by the U.\ S.\  Department of Energy under Grant No. DE-FG02-92ER40972.


\end{document}